\documentclass[epj]{webofc}
\usepackage[varg]{txfonts}   
\usepackage{float}
\usepackage{wrapfig}
\wocname{EPJ Web of Conferences}
\woctitle{INPC 2013}
\begin{document}
\title{Feasibility of the Spin-Light Polarimetry Technique for Longitudinally Polarized Electron Beams}
%
%
\vspace{-1mm}
\author{Prajwal Mohanmurthy\inst{1}\fnsep\thanks{\email{prajwal@mit.edu}} \and
        Dipangkar Dutta\inst{1}
}

\institute{Mississippi State University, MS 39762-5167, USA
}
\abstract{%
A novel polarimeter based on the asymmetry in the spacial distribution of synchrotron radiation will make for a fine addition to the existing  M{\o}ller and Compton polarimeters. The spin light polarimeter consists of a set of wiggler magnet along the beam that generate synchrotron radiation. The spacial distribution of synchrotron radiation will be measured by ionization chambers. The up-down (below and above the wiggle) spacial asymmetry in the transverse plain is used to quantify the polarization of the beam. As a part of the design process, effects of a realistic wiggler magnetic field and an extended beam size were studied. The perturbation introduced by these effects was found to be negligible. Lastly, a full fledged GEANT-4 simulation was built to study the response of the ionization chamber.
\vspace{-5mm}
}
\maketitle
\vspace{-3mm}
\section{Introduction}
\vspace{-1mm}
\label{intro}
$\>\>\>$ A 1993 proposal from Karabekov and Rossmanith explored the possibility of measuring the electron beam polarization using the synchrotron radiation produced by a magnet \cite{karabekov93}.  In this paper we examine the feasibility of a ``spin-light'' polarimetery technique for measuring longitudinal polarization of multi-GeV electron beams while building on the 1993 proposal. The simulated wiggler magnetic field was implemented in a full-fledged Geant4 simulation of the polarimeter. A polarimeter based on  spin-light would provide for a polarization measurement independent of both Compton and M{\o}ller polarimeters. A relative spin-light polarimeter could also be used in association with either a Compton or a M{\o}ller polarimeter. Highly precise, multiple independent polarimeters are a must if the ambitious goal of $\sim$ 0.5 \% uncertainity in polarimetry is to be achieved at an Electron Ion Collider (EIC) in order to meet the experimental demands.
\vspace{-3mm}
\section{Spin-Light Charecteristics}
\vspace{-1mm}
\label{sec-1}
$\>\>\>$ The spin-dependent SR distribution as given by Sokolov et. al. \cite{[4]}, ignoring higher order effects, is of particular interest as it expresses the distribution in terms of physical parameters such as spin of the electron - '$j$' and the vertical angle - $\psi$ (in the electron's frame of reference)  \cite{[5]}. The verticle angle is important as it determines the geometry of the apparatus besides other design parameters such as position of collimators.
\begin{eqnarray}
 N_{\gamma}(long) & = & \frac{9 n_e}{16 \pi^3}\frac{e^2}{cm_eR^2}\gamma^4\int_{0}^{\infty}\frac{y^2dy}{(1+\xi y)^4}\oint d\Omega (1 + \alpha ^2)^2 \times \nonumber\\
& & \left[K^2_{2/3}(z) + \frac{\alpha^2}{1 + \alpha^2}K^2_{1/3}(z) 
+ j\xi y \frac{\alpha}{\sqrt{1+\alpha^2}}K_{1/3}(z)K_{2/3}(z)\right]
\label{eq3}
\end{eqnarray}
where $ \xi  = \frac{3 B}{2 B_c} \gamma$, $B_c$ being the magnetic field under the influenze of which the entire kinetic energy of the electron is expelled as one SR photon, $y = \frac{\omega_o}{\omega_c}$, $K_n(x)$ are modified Bessel functions, $n_e$ is the number of electrons and, $z= \frac{\omega}{2\omega_C}(1 + \alpha^2)^{3/2}$, and $\alpha = \gamma \psi$. For an electron that is polarized, the power below ({\it i.e.} $-\frac{\pi}{2} \leq \psi \leq 0$) and above (i.e. $0 \leq \psi \leq \frac{\pi}{2}$) are spin dependent. More importantly the difference between the power radiated above and power radiated below, called \textit{Spin-Light} with an assymetry which can be defined as $A = \frac{\Delta N_{\gamma}}{N_{\gamma}}$, is directly spin dependent and this opens up the possibility of a direct measurement technique.

\begin{wrapfigure}{h}{0.52\textwidth}
\centering
\includegraphics[width=0.48\textwidth]{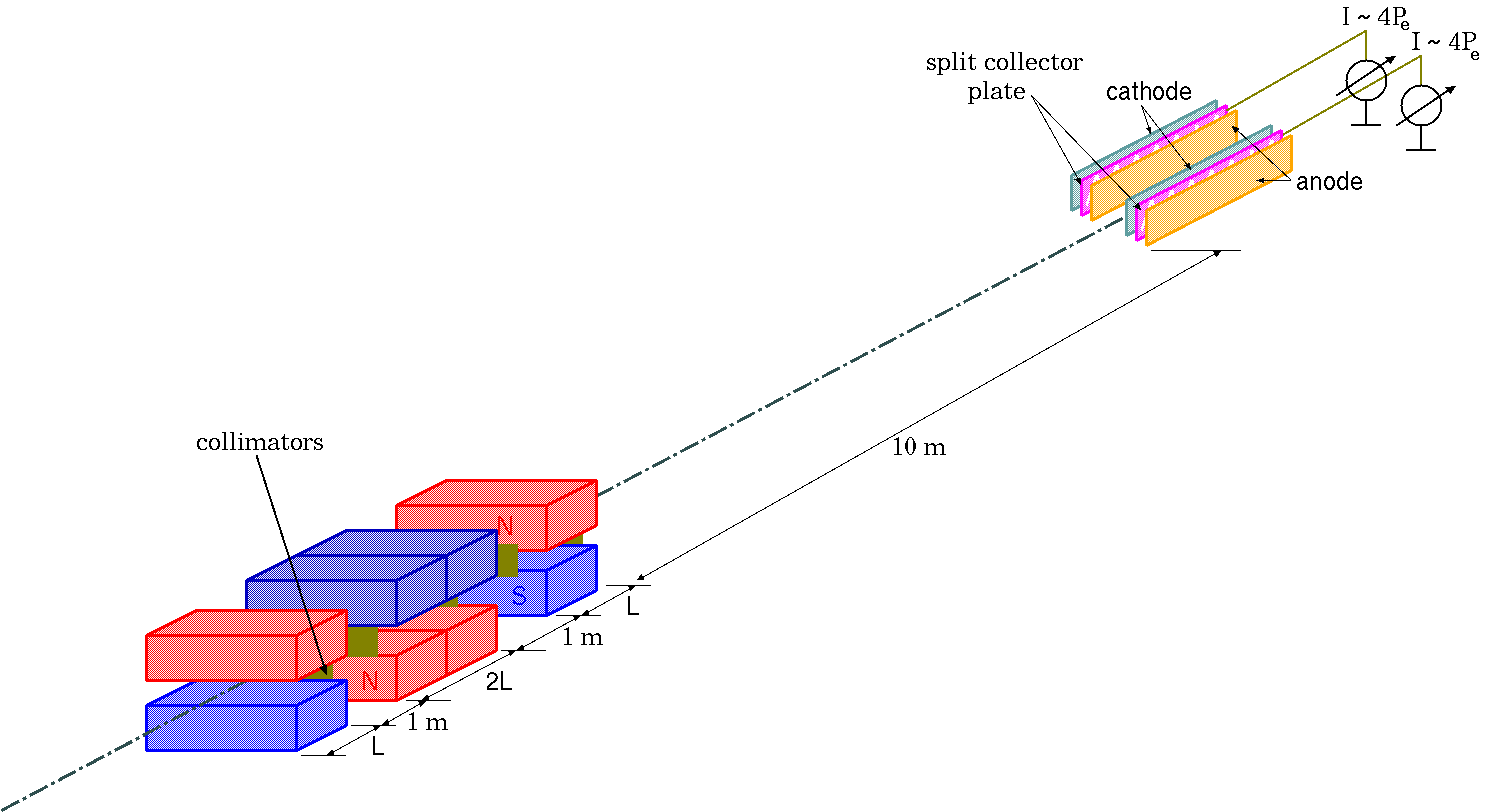}
\caption{Schematic diagram of a differential spin-light polarimeter}
\label{fig4}
\vspace{-5mm}
\end{wrapfigure}

\vspace{-3mm}
\section{Conceptual Design}
\begin{figure}[h!]
\vspace{-2mm}
\centering
    \includegraphics[scale=.415]{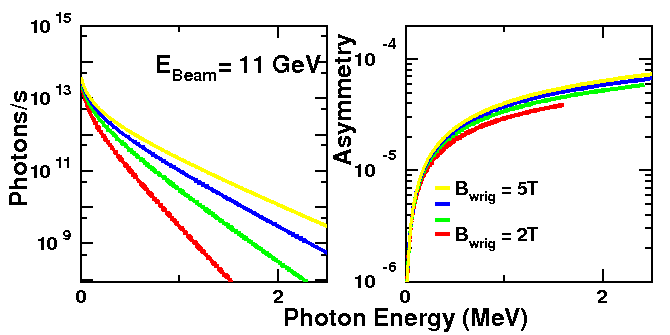}
    \includegraphics[scale=.3]{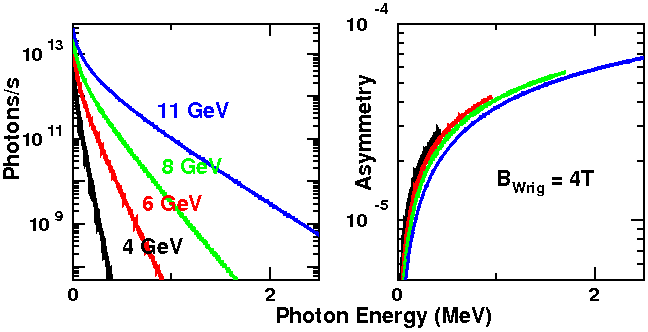}
\caption[]{(Left- Right): A. Plot of spin light spectra for various pole strengths from $2T - 5T$; B. Plot of asymmetry vs. photon energy for various pole strengths.; C. Plot of spin light spectra for various beam energies ranging from $4GeV - 12 GeV$.; D. Plot of Asymmetry vs. the photon energy for various beam energies.}
\label{figsra}
\vspace{-5mm}
\end{figure}

\vspace{-1mm}
\label{sec-2}
$\>\>\>$ The wiggler magnet is at the heart of the setup where the SR photons are produced and the ionization chambers can be used to characterize the SR in order to measure the asymmetry. It is important to note the presence of collimators on the faces of the wiggler magnets in order to prevent intermixing of the SR light  fans. Collimation creates four SR spots, with each ionization chamber receiving two collimated SR spots. Two major variables in this setup are the electron beam energy and the wiggler pole strength.  In Figures \ref{figsra}(A) \& (B), Spin-Light spectra and the asymmetry are plotted for  various wiggler pole strengths with a $11GeV$ beam and in Figures \ref{figsra}(C) \& (D), the same are plotted for various beam energies with a $4T$ wiggler field. A numerical integration code is used to generate the SR spectra and asymmetry using Eq.(\ref{eq3}).

\vspace{-3mm}
\subsection{Effects of realistic dipole magnetic field with fringes}
\begin{figure}[h!]
\centering
    \includegraphics[scale=.35]{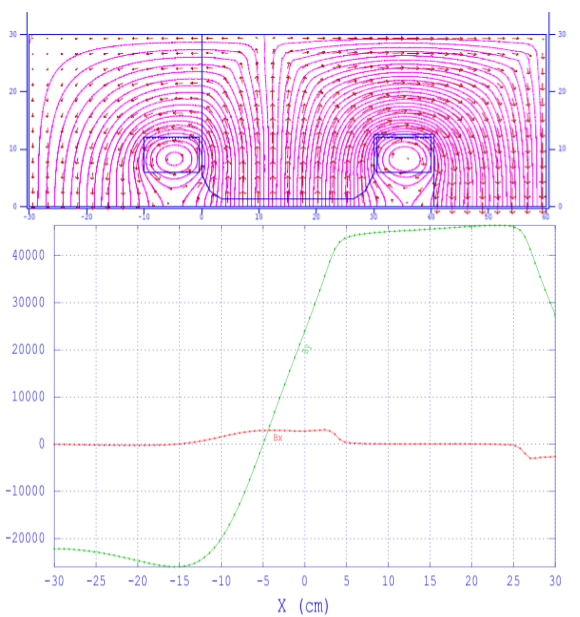}
    \includegraphics[scale=0.32]{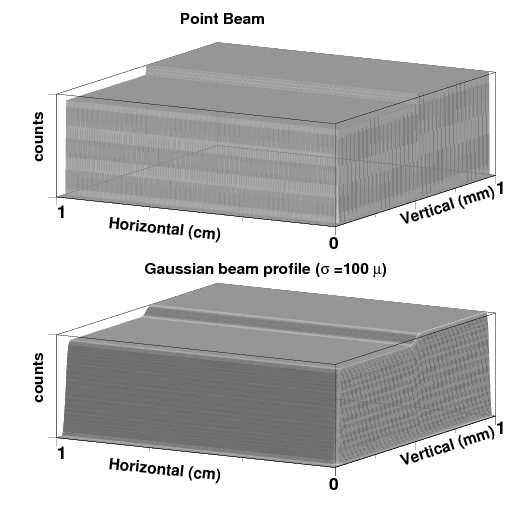}
\caption[]{(Anticlockwise from Top-Left): A. Field map of the dipole face at the edge of the dipole.; B. Plot of both the $x$ and $y$ components of the magnetic field on the transverse plane at the the edge of the dipole (Beam pipe is centered around $15$cm mark along the 'x' axis).;C. Integrated power spectra of SR Light at the IC due to Gaussian beam. (The difference between the profile has been enlarged for clarity); D. Integrated power spectra of SR Light at the IC due to a point beam.}
\label{figmapp2}
\vspace{-7mm}
\end{figure}

\begin{wrapfigure}{h!}{0.5\textwidth}
\vspace{-3mm}
\centering
    \includegraphics[scale=.48]{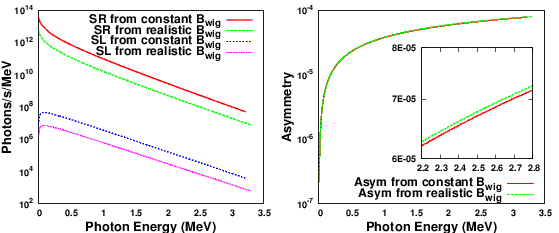}
\caption[]{(Left- Right): A. Plot showing the SR - Light and Spin - Light power spectra with a realistic wiggler magnetic field (Power spectra for uniform magnetic field have also been presented as .; B. Plot of the assymetry with a realistic wiggler magnetic field.}
\label{fignewsra}
\vspace{-7mm}
\end{wrapfigure}

\vspace{-1mm}
$\>\>\>$ A field map of the wiggler magnets can be generated by solving Maxwell's equations with appropriate boundary conditions. In \textit{LANL Poisson SuperFish} \cite{[18]}, the magnet contours can be easily defined. The field map of the magnet can then be plotted. Here, the field map at the edge where the electron beam enters the magnet is presented in Figure \ref{figmapp2}(A). Note that the beam pipe is going at the center below the magnet pole. In Figure \ref{figmapp2}(A), the physical taper of the cores can be noticed, since it is at the edge of the magnet face. The singularities seen in Figure \ref{figmapp2}(A) are the areas where the current cuts the plane. Also, it is important to note that the entire `C' magnet is not visible in the field-map,  only one half of the `C' magnet is shown in the field map. The field map obtained can be used in the numerical integration code, in place of a constant pole strength, to plot the SR spectra and the asymmetry which are presented in Figures \ref{fignewsra}(A) \& (B). Even though there is a reduction in the total power output of SR light by introducing a realistic dipole field, the asymmetry has not changed. This implies that the changes introduces by the realistic dipoles are negligible.

\vspace{-3mm}
\subsection{Effects of Extended Beam Size}
\vspace{-1mm}
$\>\>\>$ The effect of having an extended beam size of about $100\mu m$ was studied by essentially superimposing the SR Spectra generated by each differential element of the beams weighted with Gaussian distribution in order to make the extended beam a perfect Gaussian beam. The cumulative spectra for a point beam was obtained by setting the weighting factor to one. The cumulative spectra for a Gaussian beam when plotted has approximately the same structure as the spectra for the point - cross section beam. This is so because the size of the beam ($R_{beam} = 100 \mu m$) is small compared to the size of the collimated SR - Light spot which is about $1mm$ big. For the beam with a point cross section, the SR - profile is rather 'box' like at the ionization chamber. When an extended beam, that is of Gaussian profile, is introduced, the SR - profile gets a taper which is Gaussian in nature too as illustrated in Figures \ref{figmapp2}(C) \& (D). 

\vspace{-3mm}
\section{Spin Light in Geant4 and SR-Spectra}
\begin{wrapfigure}{h}{0.34\textwidth}
\vspace{-15mm}
\centering
\includegraphics[scale=.27]{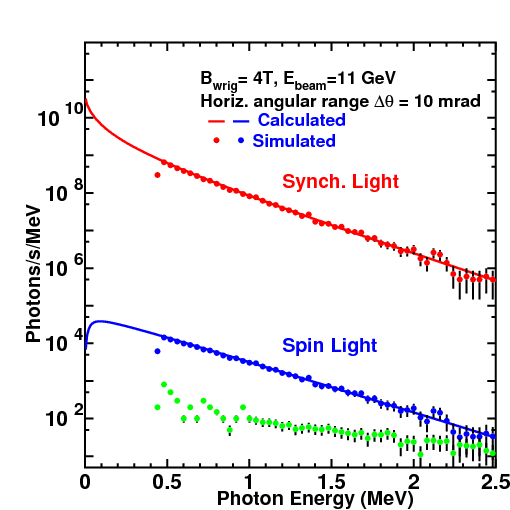}
\caption{Geant4 SR \& SL spectra as compared with physics SR \& SL spectra.}
\label{fig37}
\vspace{-10mm}
\end{wrapfigure}

\vspace{-1mm}
$\>\>\>$ A Geant4 simulation with a geometry as in Figure \ref{fig4} using EM Extra process list was constructed. Even though the integrated energy spectrum is validated in Geant4, it does not contain the angular dependence of SR light. We implement the angular dependence of SR light at the stacking action level. SR photons are killed with a probability equal to asymmetry which can be calculated from standard Geant4 track parameters. This creates an up-down asymmetry in the SR cones which is vital for this simulation of a Spin-Light polarimeter. As seen in Figure \ref{fig37}, the simulated SR-Spectra closely matches the physics SR-Spectra within $1\%$. Spin light component is obtained by subtracting the remaining tracks with positive momentum (corresponding to $0 \leq \psi \leq \frac{\pi}{2}$)  from tracks with negative momentum (corresponding to $-\frac{\pi}{2} \leq \psi \leq 0$). Figure \ref{fig37} also shows non-SR events in green which are significantly small in number compared to the corresponding spin-light events in a particular photon energy bin.
\vspace{-4mm}
\section{Conclusion}
\vspace{-1mm}
$\>\>\>$ The figure of merit for such a polarimeter increases with electron beam energy and the strength of magnetic field used. On the other hand, the SR profile becomes more compact with increase in electron beam energy. This makes it difficult for an IC to charecterize the SR profile given that the SR load increases as fourth power of beam energy. A 3 pole wiggler with a field strength of $4 T$ and a pole length of $10 cm$ would be adequate for such a polarimeter. Locating a reasonable piece of beam-line real estate is however very challenging. Given that the eRHIC design of EIC involves using a Gatalin gun \cite{[25]} at a very high rate, the recovery time of the spin-light IC will need to be extremely small if every bunch of the electron beam is to be measured for polarization, a goal which is nearly impossible with this design. A Spin-light polarimeter is apt for measuring averaged polarization of a number of beam bunches. A survey of all experiments beings proposed and their corresponding polarimetry requirements both in terms of precision of polarimetry required and the rate of measurement will go a long way in helping pin down the instrument specifications.

\end{document}